\documentstyle[11pt,epsfig]{article}

\columnwidth\textwidth

\newcommand{\avrho}{\mbox{$\langle \rho\rangle$}}
\newcommand{\avrhot}{\mbox{$\langle \rho(t)\rangle$}}
\newcommand{\avrhop}{\mbox{$\langle\rho'\rangle$}}
\newcommand{\avrhopt}{\mbox{$\langle\rho'(t)\rangle$}}

\def\gsim{\;
\raise0.3ex\hbox{$>$\kern-0.75em\raise-1.1ex\hbox{$\sim$}}\;
}
\def\lsim{\;
\raise0.3ex\hbox{$<$\kern-0.75em\raise-1.1ex\hbox{$\sim$}}\;
}

\begin{document}
\begin{flushright}
FTUV-9830\\
IFIC-9830\\
hep-ph/9809376
\end{flushright}

\begin{center}

\Large{\bf  Neutrino Conversions in Solar Random Magnetic Fields.}
\end{center}

\begin{center}
V.B Semikoz$^{a,b}$, E. Torrente-Lujan$^b$. \\
e-mail: torrente,semikoz@flamenco.ific.uv.es.\\
\end{center}

\begin{center}
{\it $^a$ The Institute of the Terrestrial Magnetism, the
Ionosphere and
Radio Wave Propagation of the Russian Academy of Sciences\\
IZMIRAN,Troitsk, Moscow region, 142092, Russia}
\end{center}
\begin{center}
{\it $^b$ Instituto de F\'{\i}sica Corpuscular - C.S.I.C.,
Departament de F\'{\i}sica Te\`orica\\ Universitat de Val\`encia,
46100 Burjassot, Val\`encia, Spain.}
\end{center}

\begin{abstract}
We consider the effect of a random magnetic field in the 
convective zone of the Sun superimposed to a regular magnetic field 
 on resonant neutrino spin-flavour oscillations.
We argue for the existence of a field of
 strongly chaotic nature at the bottom of the convective zone.
In contrast to previous attempts
we employ  a model motivated 
regular magnetic field profile: it is a static field, solution
to the solar equilibrium hydromagnetic equations. These solutions
has been known for a long time in the literature, 
we show for
the first time that in addition they are twisting solutions.
In this scenario
 electron antineutrinos are produced 
through  cascades like $\nu_{eL}\to \nu_{\mu L}\to \tilde{\nu}_{eR}$, 
The detection of
$\tilde{\nu}_{eR}$ at Earth
 would be a long-awaited signature of 
the Majorana nature of neutrino.   
The expected signals in the different experiments (SK,GALLEX-SAGE,Homestake)
 are obtained as a function of the level of noise,
 regular  magnetic field and neutrino mixing parameters. 
Previous results obtained for small mixing and 
 ad-hoc  regular magnetic profiles are reobtained.
We confirm  the strong suppression for large part 
of the parameter space of the $\tilde{\nu}_{eR}$-flux 
for high energy boron neutrinos in agreement with present 
data of the SK experiment. 

We find that MSW regions ($\Delta m^2\approx 10^{-5}$ eV$^2$, both small 
and large mixing solutions) are stable up to very large levels 
of noise (P=0.7-0.8) but they are acceptable from the point of view of antineutrino production only for moderate levels of noise ($P\approx 0.95$).

For strong noise and reasonable regular magnetic 
field, any parameter region 
$(\Delta m^2, \sin^2 2\theta)$ is excluded. 
As a consequence, we are allowed  to 
reverse the problem and to put limits on r.m.s field strength  and 
transition magnetic moments by demanding a particle 
physics solution to the SNP  under this scenario.

\end{abstract}


\newpage

\section{Introduction}
A neutrino transition magnetic moment can account, 
both for the observed 
deficiency of the solar neutrino flux and the time variations of the signal. 
The overall deficit is caused by the alteration or suppression 
of the neutrino energy 
spectrum.
The time dependence may be caused by time variations of the magnetic 
field in the convective zone of the Sun.
As it has been shown in \cite{akh4}, magnetic moments solutions are not ruled 
out by present experimental data; that is valid for both, absolute deficits 
and time variations of the observed solar neutrino flux.

In transverse magnetic fields, neutrinos with transition magnetic moments will 
experience spin and flavour rotation simultaneously (resonant 
spin-flavour precession, 
RSFP). 
The observation of electron antineutrinos from the Sun would lead to the conclusion that the 
neutrinos are Majorana particles. There are however stringent bounds on the presence of solar 
electron antineutrinos 
coming from the High energy Boron neutrinos
( $E_{\nu}>\approx  8.3~MeV$ \cite{fio1,fio2}).

Magnetic fields measured on the surface of the Sun are
 weaker than within 
the
interior of the 
convective zone. The mean field value over the solar disc is  of the order of 1 G and in the 
solar spots magnetic field strength reaches $\ 1$ KG. In magnetic 
hydrodynamics (MHD) one can explain
such fields in a self-consistent way if these fields are generated by dynamo mechanism at the 
bottom of the convective zone.
In this region  the strength of small scale 
 regular magnetic fields could  reach a value of 
100 KG.
These fields propagate through the  convective zone and photosphere 
decreasing in the strength value 
while increasing in the scale giving traces in from of loops in bipolar
 active regions (solar spots).

Large-scale toroidal magnetic field created by dynamo mechanism in
 convective zone 
has strength even less than small-scale r.m.s fields near the bottom of convective zone. This is 
the main reason why one should  consider 
 neutrino propagation in the random magnetic field of the Sun. 
The ratio of the r.m.s. random field and the regular (toroidal) 
field may be $\sim$ 20-50, therefore the 
problem of RSFP neutrino propagation  in noisy magnetic field seems to be important. 

Estimations for the ratio of rms fields to regular field  are necessarily very 
rough.
In textbooks \cite{vai1,vai2,park1,park2}  
we find the conservative ratio 
$$\langle\tilde{B}^2\rangle/B^2_0\sim 1.$$ 
In more elaborate models  the ratio of magnetic energy densities is given by the magnetic 
Reynolds number, 
$$\langle\tilde{B}^2\rangle/B^2_0\sim R_m^{\sigma},$$ 
which may be  much bigger than unity for plasma
 with large conductivity. Here 
$\sigma>0$ is a topology index \cite{vai1,vai2,Cataneo,Diamond}.

The effect of random magnetic fields in RSFP solutions to the SNP and 
antineutrino production has been explored  previously for simplified
 models \cite{tor3,tor4}.
In this work we will deal with the complete problem, 
we will present calculations of neutrino spin flavour conversions in 
presence of matter and magnetic field. 
The magnetic field will have two ingredients. The first ingredient will be a 
  a theoretically motivated solar
magnetic field profile.
This magnetic field, which is one  solution to the  static 
magnetic hydrodynamic equations, is in addition a twisting field. 
The degree of twisting, the
 transversal profile and the ratio core field/ convection field 
are functionally related under this model.
Of all possible solutions, 
in practical calculations we have considered only the solution with minimum
twist  which is at the same time the solution which implies a minimum 
value for the central field ( a factor $2$ or less  of the magnetic field 
at the convective zone).
As a second ingredient,
 the effect of a layer of magnetic noise generated at the bottom of the convective zone has been included, we have justified that the level of noise
in this region can be certainly very high.

Notice that an analogous problem but without the regular twist field
motivated here while with use of a direct numerical method for dealing with
the random magnetic fields in the solar convective zone has been 
recently considered (\cite{bykov}). The results obtained in \cite{bykov} 
are in general  consistent with those obtained in the present work (see 
 Conclusions below).

In Section (2) we present the effective Hamiltonian  
governing the evolution of the 
4x4 neutrino wave function (two flavors times two helicities) in presence of a generic magnetic field. 
In Section (3.1) we give some physical 
arguments supporting the idea that r.m.s fields could be 
indeed important in the solar convective region.
The analytical expressions for the hydrodynamic-inspired  regular large-scale magnetic field appear
 in Section (3.2). There we show how different 
configurations of this field can be classified according to 
its intrinsic ''twist'' in the perpendicular plane to 
neutrino propagation.
In Section (4) we give the master equation governing the 
time evolution of the averaged density matrix in presence of
the  random  magnetic field. We show analytically how in the
limit where chaotic field dominates, the 4x4 evolution 
equations decouples and the 
averaged transition probabilities 
follow a Markovian process.  
The results of the numerical integration of the exact 
averaged master equation are shown and discussed 
in Sections (5) and (6). From the 
calculated transition probabilities 
the   expected total 
signals in each of the existing experiments
(SK, SAGE-GALLEX,HOMESTAKE) are obtained and 
$\chi^2$-allowed parameter regions. In addition the 
expected electron antineutrino signal in SK is calculated and
compared with existing bounds.

\section{The Master equation}

There are two channels for $\nu_{eL}\to \nu_{eR}$-conversions 
corresponding to the cascades:
(i) $\nu_{eL}\to \nu_{\mu L}\to \tilde{\nu}_{eR}$ or (ii)
$\nu_{eL}\to \tilde{\nu}_{\mu R}\to \tilde{\nu}_{eR}$. 
The former case realizes if 
we assume a
zero field in  the radiative 
zone and in the core or if 
we exclude right-handed neutrino
production there and the MSW-conversion ($\nu_{eL}\to \nu_{\mu L}$) takes place before
neutrino reaches convective zone.
 The latter case (ii) realizes when a strong magnetic field is present both 
in the radiative zone
and in the solar core. 
In this case the RSFP takes place before the MSW-conversion since
the $\nu_{eL}-\tilde{\nu}_{\mu R}$ energy splitting $V = G_F\sqrt{2}(N_e(r_1) - N_n(r_1))$ is 
less than for the MSW one ($\nu_{eL}\to \nu_{\mu l}$) at the same point $V = 
G_F\sqrt{2}N_e(r_1)$. This is true for 
the typical  admixture 
of helium and heavy 
elements considered to be present in the inner layers of the Sun.

We consider conversions $\nu_{eL,R}\to \nu_{aL,R}$, $a = \mu$ or
$\tau$, (for definiteness we will refer to  $\mu$ for the rest of this work) for two neutrino flavors obeying the 
master evolution equation
\begin{eqnarray}
i \partial_t 
\pmatrix{\nu_{eL}\cr\overline{\nu}_{eR}\cr\nu_{\mu L}\cr\overline{\nu}_{\mu R}} 
&= &
\pmatrix{V_e -c_2\delta &0&s_2\delta &\mu B_{\perp}^+(t)\cr 
0& - V_e - c_2\delta & - \mu B_{\perp}^-(t)& s_2\delta \cr
s_2\delta & - \mu B_{\perp}^+(t)&V_{\mu} + c_2\delta &0\cr
\mu B_{\perp}^-(t)&s_2\delta &0& - V_{\mu} + c_2\delta}
\pmatrix{\nu_{eL}\cr\overline{\nu}_{eR}\cr\nu_{\mu L}\cr\overline{\nu}_{\mu R}} 
\label{e1001}
\label{master}
\end{eqnarray} 
where $c_2 = \cos 2\theta$, $s_2 = \sin (2\theta)$, $\delta = \Delta m^2/4E$ are the neutrino 
mixing parameters; $\mu = \mu_{12}$ is the neutrino active-active
transition magnetic moment; 
\begin{eqnarray}
B_{\perp}^\pm(t)& =& B_{0\perp}^\pm(t) + \tilde{B}_{\perp}^\pm(t)
\label{e7701}
\end{eqnarray}
 is the magnetic field component which is perpendicular to the neutrino trajectory in the Sun; 
$$V_e(t) = G_F\sqrt{2}(\rho (t)/m_p)(Y_e - Y_n/2)$$ 
and 
$$V_{\mu}(t) = G_F\sqrt{2}(\rho (t)/m_p)(- Y_n/2)$$ 
are the neutrino 
vector potentials for $\nu_{eL}$ and $\nu_{\mu L}$ in the Sun given 
by the abundances of the electron ($Y_e = m_pN_e(t)/\rho (t)$) and 
neutron ($Y_n = m_pN_n(t)/\rho(t)$) components and 
by the SSM density profile \cite{BP95}
\begin{eqnarray}
\rho (t) &=&250~gcm^{-3}\exp ( - 10.54\ t).
\label{e7703}
\end{eqnarray}

The transverse magnetic field  $B_{\perp}(t)$ 
appearing in Eq.(\ref{e7701}) is 
given by the following expression where Cartesian and polar coordinates 
are written explicitly:
\begin{equation}
 B_{\perp}^{\pm}
(t) \equiv B_x(t) \pm i B_y(t)\equiv |B_{\perp}(t)|e^{\pm i\Phi (t)}
\label{e9001}
\end{equation}
where 
\begin{equation}
\Phi (t) = \arctan B_y/B_x.
\label{e9002} 
\end{equation}
The Cartesian form is useful in writing the averaged master equation. The 
polar form is the most convenient in separating out the twisting part.

The nature and magnitude for the 
regular  ($B_{0\perp}$) and chaotic parts ($\tilde{B}_\perp$) of the 
magnetic field will be the subject of the next section. The equation for 
the evolution of the average density matrix corresponding to the master 
equation (\ref{e1001}) can be found using the formalism 
developed in  \cite{tor2} and will presented later in this this work.

The magnetic field strength enters the evolution Eq.(\ref{e1001}) being multiplied 
by the neutrino transition moment $\mu$. The existing upper limits on the magnetic moment of the electron neutrino include the laboratory bound
$\mu< 3-4\times 10^{-10}\mu_B$  from reactor experiments as  well as stronger (one or two orders or magnitude) astrophysical and cosmological limits. 
In our calculations we will consider always  the product $\mu B$. 
Expected values of $B\approx 1-100$ kG in the Sun convective zone and 
$\mu=10^{-11} \mu_B$ would give an expected range for the 
 product $\mu B\approx 10^{-8}-10^{-6} \ \mu_B G\approx 5.6\ 10^{-17}-10^{-15} $ eV  or in the 
practical units which will be used 
 throughout
 this work $\mu B\approx \ 0.1-10.0\ \mu_{11} B_4$.

\section{Solar magnetic fields}

\subsection{Random magnetic fields}

The r.m.s. random component $\sqrt{\langle \tilde{B}^2(t)\rangle}$ 
can be comparable with the regular one, $B_0(t)$, and maybe even
much stronger than $B_0$, 
if  a large magnetic Reynolds number $R_m$ leads to the effective dynamo 
enhancement of small-scale (random) magnetic fields.

Let us give simple estimates of the magnetic Reynolds number $R_m = lv/\nu_m$
in the convective zone for fully ionized hydrogen plasma ($T\gg I_H\sim 13.6~eV\sim 10^5~K$). 
Here $l\sim 10^8~cm$ is the size of eddy (of the order of magnitude of a granule size) with the 
turbulent velocity inside of it
$v\sim v_A\sim 10^5~cm/s$ where $v_A = B_0/\sqrt{4\pi \rho}$ is the Alfven velocity for MHD 
plasma,
$B_0$ is a large-scale field in convective zone and $\rho$ is the matter density (in $g/cm^3$) in 
the SSM.

The magnetic diffusion coefficient $\nu_m = c^2/4\pi \sigma_{cond}$  
($\sim$ magnetic viscosity) enters the diffusion term  of 
the Faraday equation,
\begin{equation}
\frac{\partial \vec{B}(t)}{\partial t} = rot [\vec{v}\times \vec{B}(t)] + \nu_m \Delta \vec{B}~.
\label{Faradey}
\end{equation} 
Here $c$ is the light velocity; 
the conductivity of the hydrogen plasma $\sigma_{cond} = \omega_{pl}^2/4\pi 
\nu_{ep}$. 
$$\omega_{pl} = \sqrt{4\pi e^2n_e/m_e} = 5.65\times 10^4\sqrt{n_e}~s^{-1}$$ 
is the plasma (Langmuir) frequency;
$\nu_{ep} = 50n_e/T^{3/2}~s^{-1}$ is the electron-proton collision frequency, 
the electron 
density $n_e( = n_p)$ ($cm^{-3}$)
and the temperature $T$ (K).

Thus we find that the magnetic diffusion coefficient 
$$\nu_m\simeq 10^{13}(T/1~K)^{-3/2}\ {\rm cm}^2{\rm s}^{-1}$$
 does not depend 
on the charge density $n_e$ and it is very small (the Reynolds number is big) for hot plasma 
$T\geq 10^5~K\gg 1~K$. Actually, from the 
comparison of the first and second terms in the r.h.s. of the 
Faraday equation Eq. (\ref{Faradey}) we find that $v/l\gg \nu_m/l^2$,
 or $\nu_m\ll vl\sim 10^{13}~cm^2s^{-1}$ since $T/1~K\gg 1$.
This means that the magnetic field in the Sun is mainly {\it frozen-in}. 
Neglecting 
the second term in Eq. (\ref{Faradey}) and using the Maxwell equation 
$$rot \vec{E}= - c^{-1}(\partial \vec{B}/\partial t)$$ 
we obtain the condition for frozen-in field:
the Lorentz force vanishes, 
$$\sim (\vec {E} + [\vec{v}\times \vec{B}]/c)\approx 0$$ 
but the current 
$$\vec{j} = \sigma_{cond}(\vec {E} + [\vec{v}\times \vec{B}]/c)$$
remains finite if the conductivity is large, $\sigma_{cond}\to \infty$.

The magnetic Reynolds number 
$$R_m = lv\omega_{pl}^2/(c^2\nu_{ep})\simeq lv\times 10^{-13}(T/1~K)^{3/2}~cm^2s^{-1}$$ 
is huge if we substitute the estimate
$lv\sim 10^{13}~cm^2s^{-1}$ given above. A large value for the 
Reynolds number  is a necessary condition for an effective  
dynamo enhancement in the convective zone.

The estimation of the quantity $\eta$ for the solar 
convective zone (and other cosmic dynamos) is the matter 
of current scientific discussions.
The most conservative estimate, simply based on
 equipartition, is $\eta=constant$. 
According to  direct observations of galactic
magnetic field presumably driven by a dynamo, 
$\eta \approx 1.8 $ \cite{Ruzmaikin}. 
A more developed theory of equipartition
gives, say, $\eta \approx 4 \pi  \ln R_m$ 
(see \cite{Zeldovich}). 
Notice that this estimate is considered now as
very conservative. 
Basing on more detailed theories of MHD turbulence
estimates like $b \sim \sqrt{R_m} B$
are discussed \cite{Cataneo}, \cite{Diamond}.

The 
random magnetic field component in the 
Sun ($\langle \tilde{B}(t)\rangle = 0$) will be  described in general by 
an arbitrary correlator 
$$\langle \tilde{B}(t)\tilde{B}(t^{'})\rangle = 
\langle \tilde {B}^2\rangle f(t - t^{'}).$$
 We will assume that 
the strength of
the r.m.s. field squared $\langle \tilde{B}^2\rangle = \eta B_0^2$ 
is
parametrized by the 
dimensionless parameter $\eta = R_m^{\sigma}>1$ , which it  can be, in general, much bigger than unity, 
$\eta\gg 1$. 

The correlator function $f(t)$ is unknown a priori but it
 takes the particular
$\delta$-correlator form
$f(t) = L_0\delta (t)$ 
if the correlation length (for two neighboring magnetic field domains) is much less than the 
neutrino oscillation length, $L_0\ll l_{osc}$.   
$L_0$ can be considered  a free parameter
 ranging in the interval $1-10^4$ km.
In the averaged evolution equations it appears only the 
product  $\eta L_0$. Thus  
in what follows we will present our results as a function of the
quantity P which  is a simple function of such product:
\begin{eqnarray}
P&=&\frac{1}{2}\left (1+ \exp(-\gamma)\right )\\
\gamma&\equiv&\frac{4}{3} \Omega^2 \Delta t\equiv 
\frac{4}{3} \eta L_0 (\mu B_{0})^2 \Delta t.
\label{e2003b}
\end{eqnarray}

\begin{figure}[p]
\centering\hspace{0.8cm}
\begin{tabular}{c}
\epsfig{file=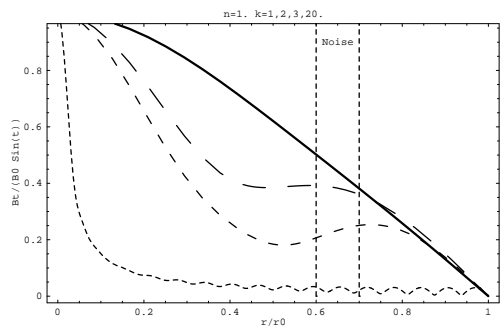,height=9cm}
\end{tabular}
\caption{Transversal magnetic field profile. Solutions
$B_{0\perp}= B_0 \sin\theta f(r)/r$ (see Eq.(\protect\ref{e9004}))
with $k=1$ (Continuos line) $2,3,30 $ respectively.}
\label{f7}
\label{f6}
\end{figure}

The reason for using $P$ is that it  is a good approximation for the 
 depolarization that the presence of noise induces in the averaged 
neutrino density matrix. 
 $\Delta t$ is the distance over which the noise is acting.
We have  supposed in our computations that the noise is effective only in a thin layer with thickness 
$\Delta t=0.1\ R_\odot$ starting  at $r=0.7\ R_\odot$, the
bottom of the convective zone. This is represented together with the 
regular transverse profile in Fig.(\ref{f7}). 
 In Table (\ref{t2}) the quantities 
 $\sqrt{\langle B_0^2\rangle}$ and  $\eta$ are  computed for a given $P$ 
supposing the 
 reasonable value  $L_0=1000$ Km.

\subsection{Regular large-scale magnetic field in the Sun. Twist field.}

Many phenomenological formulas for $B_{0\perp}$ in the 
convective zone
and in the central region of the Sun  \cite{akh2}, with or without 
 twist, has been employed in the literature.
In particular, twist was applied for effective enhancement of the  
process $\nu_{eL}\to \tilde{\nu}_{eR}$ in the Sun (i.e. in Ref.\cite{bal2}). Without twist the coherent 
sum of two amplitudes for the process $\nu_{eL}\to \tilde{\nu}_{eR}$,
$$M_1(\nu_{eL}\to \nu_{\mu L}\to \tilde{\nu}_{eR}) + M_2(\nu_{eL}\to \tilde{\nu}_{\mu R}\to \tilde{\nu}_{eR}),$$
occurs proportional to the small abundance of 
neutrons and production of electron antineutrino is negligible \cite{akh4}.

In the presence of the twist, $\Phi (t)\neq 0$,
this sum becomes  proportional to the angular velocity 
$$\dot{\Phi}=\kappa R_{\odot}^{-1} = \kappa \times 0.3\times 10^{-15}~eV$$ 
where $[\kappa] = n =1,2,...$ is the effective number of revolutions 
of the field $\vec{B}_{\perp}$ in the plane which 
is perpendicular to the neutrino trajectory.
The more rounds ($n\gg 1$) take place before 
the resonant RSFP position, given by the condition
$$G_F\sqrt{2}(N_e - N_n) - 2\delta \cos 2\theta - \dot{\Phi} = 0,$$
 the deeper the resonant point happens 
(for successive (positive) sign of $\dot{\Phi}$). 
It is possible then the merging
of RSFP ($\nu_{\mu L}\to \tilde{\nu}_{eR}$)  and 
MSW ($\nu_{eL}\to \nu_{\mu L}$) resonances since the MSW resonance 
is unchanged, ($ G_F\sqrt{2}N_e 
= 2\delta \cos 2\theta$).
For the case of merging of resonances 
the  of $\nu_{eL}\to \tilde{\nu}_{eR}$-conversions 
tend to be more   
adiabatical and complete than for separated resonances \cite{akh5}.

Unfortunately, in the known MHD plasma solutions for the toroidal field 
evolving in convective zone due to dynamo mechanism 
(Yoishimura model, \cite{yoi1})
 the predicted number of revolutions for twist 
(around toroids in northern and southern hemispheres of the 
Sun) is very small:
 $n\gsim 1$. For this small  twist rate $\dot{\Phi}\sim n$ 
the resonances above are fulfilled 
for a small $\Delta m^2\sim 10^{-7}- 10^{-8}~eV^2$ and happen 
not too deep in solar interior to have important consequences.

Below we apply for the neutrino conversions described by our master equation Eq. (\ref{master}) 
the self-consistent model of large-scale regular field given in 
\cite{kut1}. 
The global solar magnetic field is the axisymmetric equilibrium solution of 
the MHD static equations (quiet Sun) in the spherically symmetric 
gravitational field of the Sun. The reasonable 
boundary condition  $B_0 = 0$ on the photosphere ( $r=R_{\odot}$) is imposed in addition.
Any field solution to these equations and boundary conditions is a twisting field 
{\it with an arbitrary small or large 
number of revolutions along radius} ($k=1,...,\infty$, 
the twist rate can be taken as label for distinguish particular solution within the family).

The spherical components of the magnetic field for the fundamental radial mode
$n = 1$ do not depend on azimuthal angle $\phi$. 
In the whole region $0\leq r\leq R_{\odot}$ they are of the form
(the different solutions are labeled by $k$):
\begin{eqnarray}
B_{0r}(r,\theta_s)& =& 2K\cos \theta_s\Bigl [1 - \frac{3R_{\odot}^2}{r^2z_{2k}\sin 2z_{2k}}\Bigl 
(\frac{\sin \alpha r}{\alpha r} - \cos \alpha r\Bigr )\Bigr ]~,\nonumber\\
B_{0\theta_s}(r,\theta_s)& =&  -K\sin \theta_s\Bigl [2 + \frac{3R_{\odot}^2}{r^2z_{2k}\sin 
2z_{2k}}\Bigl (\frac{\sin \alpha r}{\alpha r} - \cos \alpha r - \alpha r \sin \alpha r\Bigr 
)\Bigr ]~,\nonumber\\
B_{0\phi}(r,\theta_s) &=& Kz_{2k}\sin \theta_s\Bigl [\frac{r}{R_{\odot}} - 
\frac{3R_{\odot}^2}{rz_{2k}\sin 2z_{2k}}\Bigl (\frac{\sin \alpha r}{\alpha r} - \cos \alpha r\Bigr )\Bigr ]~.
\label{e9003}
\end{eqnarray}
Here $\theta_s$ is the polar angle. $z_{2k}$ is any   root of the spherical Bessel function 
$$f_n(z) = \sqrt{z}J_{n + 1/2}(z)$$
 of the first order (n =1). This equation  follows from the 
boundary condition. For the first three roots:
$$z_{2k}\alpha R_{\odot} = 5.7635,~9.0950,~12.3224,...\ (k=1,2,3,...).$$ 
The constant $K$ 
is related with
central field $B_{core}$, the only free parameter in this model:
\begin{equation}
K = \frac{B_{core}}{2(1 - \alpha R_{\odot}/\sin \alpha R_{\odot})}.
\label{central}
\end{equation}

The modulus of the perpendicular component is of the form:
\begin{eqnarray}
B_{0\perp}& =& \sqrt{B_{0\phi}^2 + B_{0\theta_s}^2} =B_{core}\frac{\sin\theta}{r} f(r)
\label{e9004}
\label{e6007}
\end{eqnarray}
where $f(r)$ is some known function of  gentle behavior.
Obviously, on the solar equator ($\theta_s = 0$) the perpendicular 
component vanishes. This is the same behavior as it is shown
 by the toroidal field in the Yoishimura model \cite{yoi1}
 \footnote{This property is general and could lead to semiannual variations of the neutrino flux 
if one takes into account inclination 
($7^o$) of the solar axis (perpendicular to solar equator) to 
the ecliptic with the Earth 
orbit.}.

Notice the regular behavior of $\vec{B}_0$ at the Sun center ($r=0)$: 
\begin{eqnarray}
B_{r}(0)&=& B_{core}\cos\theta_s,\\
B_{\theta_s}(0) &=& - B_{core}\sin \theta_s,\\ 
B_{\phi}(0)&=& B_{core}\sin \theta_sr/R_{\odot}\to 0.
\end{eqnarray}

The  twist for the perpendicular component 
$B_{0\perp}(r)$
 is defined by the angle:
\begin{eqnarray}
\Phi (r)_{z2k}& =& \arctan B_{\phi}(r)/B_{\theta}(r)\sim z_{2k}\\
\dot{\Phi} (r)_{z_{2k}}& \approx& k R_\odot^{-1}.
\end{eqnarray}
For $k=1$ for example, $\Phi$ changes between 0 and $\pi$ along the neutrino path from the
core of the Sun through the surface at an uniform rate. For higher roots 
we observe that effectively $\dot\Phi\approx k\pi$ uniformly.

According to this model, the expected magnetic field at the core is 
typically only 2-3  times (or less) 
the magnetic field at the convective zone. For the 
values  that we will consider later, $B_{0.7}\approx <100-200$ kG, the values 
corresponding at the core are well below astrophysical bounds
derived from traces of these fields at solar surface.

\section{The averaged master equation.}

The master Equation (\ref{e1001}) 
can be written in terms of the density matrix
$\rho(t)$ as:
\begin{eqnarray}
i\partial_t \rho&=&[H_{reg},\rho]+\mu \tilde{B}_x(t)[V_x,\rho]+\mu \tilde{B}_y(t)[V_y,\rho].
\label{e7702}
\end{eqnarray}
The elements
 of the matrices $H_0, V_x,V_y$ can be read off the Eq. (\ref{e1001}). 
The $\tilde{B}_x,\tilde{B}_y$ are the Cartesian transversal components
of the chaotic magnetic field. Vacuum mixing terms and matter terms corresponding to the SSM density profile given before and the regular  
magnetic part Hamiltonian, with the profile determined by the 
Eqs.(\ref{e7703}), are all included in $H_{reg}$. 
In particular, the matrices $V_{x,y}$ are given in terms of the 
Pauli matrices $\sigma_{1,2}$ by:
\begin{equation}
V_x=\pmatrix{0&i \sigma_2\cr -i \sigma_2 & 0}, \quad
V_y=\pmatrix{0&-i \sigma_1\cr  i\sigma_1 & 0}.
\end{equation}

It is our objective in this section to write the differential evolution equation for the average density matrix $\langle\rho\rangle$.


We assume that the components 
 $\tilde{B}_x,\tilde{B}_y$ are statically independent, each of them 
characterized by a delta-correlation function:
\begin{eqnarray}
\langle \tilde{B}_{x,y}(t)\tilde{B}_{x,y}(t')\rangle &=& \langle \tilde {B}_{x,y}^2\rangle  L_0\delta(t - t^{'})\\
\langle \tilde{B}_{x}(t)\tilde{B}_{y}(t')\rangle &=& 0 .
\end{eqnarray}
From now on we will make the following assumption, which is reasonable from  equipartition arguments,
$$\langle\tilde{B}_x^2\rangle=\langle\tilde{B}_y^2\rangle=\langle\tilde{B}_{0\perp}^2\rangle/2=\langle\tilde{B}_0^2\rangle/3$$

The averaged evolution equation is a simple generalization 
(see Ref. \cite{tor2} for a complete derivation) 
of the well known Redfield equation
\cite{balantekin,tor20}  for two independent sources of noise 
and reads
($\Omega^2\equiv  L_0\mu^2 \langle\tilde{B}^2\rangle/2\equiv \eta L_0 
(\mu B_{0})^2/3$):
\begin{eqnarray}
i\partial_t \avrho=[H_{reg},\avrho]-i \Omega^2[V_x,[V_x,\avrho]]-i\Omega^2 
[V_y,[V_y,\avrho]].
\label{e8690}
\end{eqnarray}

It is possible to write the Eq.(\ref{e8690}) in a more evolved form. Taking into account 
 the particular form of the matrices 
$V_{x,y}$ we can simplify 
considerably the double commutators:
\begin{eqnarray}
i\partial_t \avrho=[H_{reg},\avrho]-2 i \Omega^2\left (
2 \avrho-V_x\avrho V_x-V_y\avrho V_y \right ).
\label{e8691}
\end{eqnarray}

The second term in Eq.(\ref{e8691}), which gives the 
leading noise behavior, can be eliminated by a rescaling of the
density matrix: 
$$\avrhot=\exp(-4 \Omega^2  t)\avrhopt.$$
 With this redefinition the evolution equation reads:
\begin{eqnarray}
i\partial_t \avrhop=[H_{reg},\avrhop]+i 2\Omega^2 \left (V_x\avrhop V_x+V_y\avrhop V_y\right ).
\label{e8692}
\end{eqnarray}

The expression in Eq.(\ref{e8692}) is the one which 
 has been used in the numerical calculations to be 
presented below.
It is useful however to consider the 
solution to  Eq.(\ref{e8692})  when $H_{reg}\equiv0$. 
This is the appropriate limit when dealing with extremely low $\Delta m^2$ or 
very large energies, 
for an extreme level of noise or when the 
distance over which the noise is acting is small enough to consider the evolution given by 
$H_{reg}$ negligible.
 In any other scenario it can give at least an idea 
of the general
behavior of the solutions to the full  Eq.(\ref{e8692}). 
When $H_0=0$   only the two last terms   in the equation  remain 
and an
exact simple expression is obtainable by ordinary algebraic methods.
The full 4x4 Hamiltonian decouples in 2x2 blocks.
The quantities of interest, the averaged transition probabilities,
are given by the diagonal elements of $\avrho$. 
If $P_{f,i}$ are the final  and initial probabilities 
( at the exit and at the entrance of  the noise  region) 
their   averaged counterparts  fulfill 
linear relations among them, schematically:
\begin{eqnarray}
Q_f^{A,B}&=&M Q_i^{A,B}
\end{eqnarray}
with $Q^{A,B}$ any of the two dimensional vectors
\begin{equation}
Q^A=\pmatrix{\langle P(\nu_{eL}\to  \nu_{eL}  )\rangle \cr 
\langle P(\nu_{eL}\to \tilde{\nu}_{\mu R} )\rangle } \quad
Q^B=\pmatrix{\langle P(\nu_{eL}\to  \tilde{\nu}_{eR}  )\rangle \cr 
\langle P(\nu_{eL}\to \nu_{\mu L} )\rangle } \quad
\end{equation}
and the Markovian matrix $M$:
\begin{eqnarray}
M&=&\pmatrix{P & 1-P \cr 1-P & P}
\end{eqnarray}
with $P$ defined in Eq.(\ref{e2003b}).
It can be shown that in this simple case $P$ is exactly the final polarization of the 
density matrix. In the general case with a finite $H_{reg}$ it can be shown numerically 
that the quantity $P$ still gives a reasonable approximation ($< 10\%$) to the real polarization, at least for the 
cases of interest in this work. 

\section{Results and Discussion. }

The present status of the Solar neutrino problem represented by the ratios between the solar neutrino observations and the solar model predictions is summarized in Table (1) of Ref.(\cite{bah2}) which for 
the sake of completeness we reproduce in (\ref{t1}).
It is claimed in addition by a variety of analysis 
(check  Ref.\cite{akh10} for an useful review) that Homestake,
the chlorine radiochemical experiment,  presents an anticorrelation with the solar
activity. 
Signal time variations  are not observed by 
the Kamiokande experiment. 
The allowed time variation 
is restricted by  to be $\approx < 30\%$ at $90\%$ C.L. \cite{suz1,bah2}.

We have calculated the expected neutrino signals in the Homestake, Ga-Ge and (Super)-Kamiokande experiments. For this objective, the time averaged 
survival and transition
probabilities have been obtained 
by numerical integration of  the ensemble averaged master equation (\ref{e7702}) 
for a regular magnetic profile $B_{0\perp}$ as given
 in Eqs.(\ref{e9003},\ref{e9004}) for $k=1$
and the matter density given by Eq.(\ref{e7703}).

The   free parameters of our model are four: 
the squared mass differences ($\delta=\Delta m^2/2E$),
 the flavour mixing angle ($s_2^2\equiv \sin^2 2\theta$),  
a noise strength parameter ($P$) and 
the product of magnetic field and moment ($\mu B_{0\perp}$ 
at a given 
radius and latitude).
We have found convenient to use the
 magnitude of the transverse magnetic field 
at the bottom of the convective zone ($r=0.7 R_\odot$) for a solar latitude
$\theta_s=7^o$; 
 for other latitudes they have to be rescaled accordingly.

\begin{figure}[p]
\centering\hspace{0.8cm}
\epsfig{file=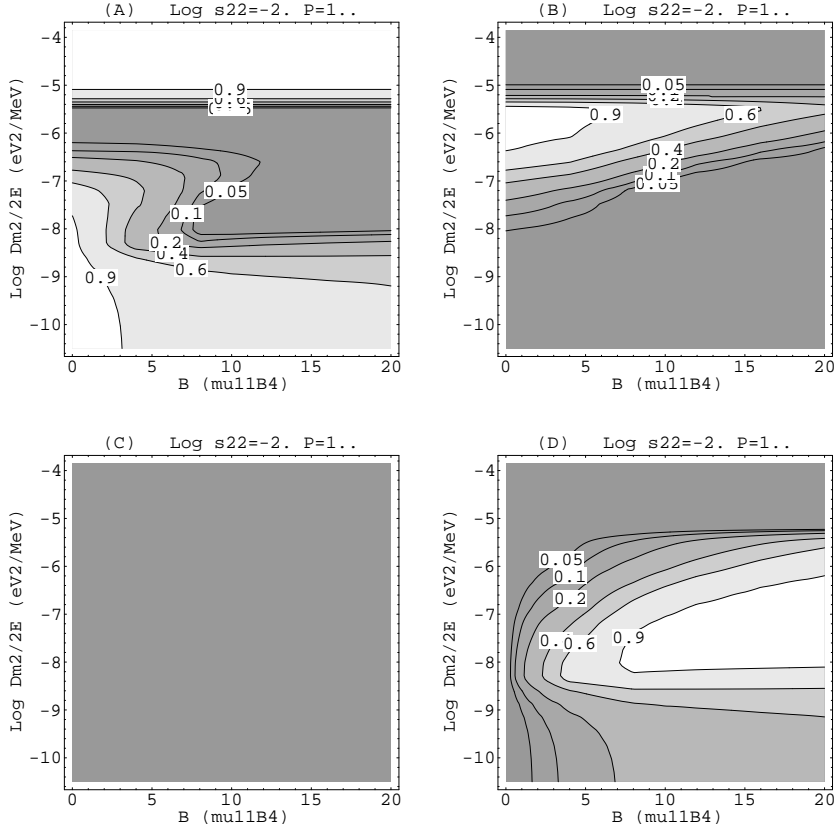,height=16cm}
\caption{Probabilities at Earth for a neutrino created at the solar center 
as a function of the magnetic field taken at the bottom of the convective zone ($B(0.7 r_0)$), for a fixed mixing angle ($s_2^2=0.01$).
Respectively A) $P_{\nu_e\nu_e}$ , B) $P_{\nu_e\nu_\mu}$, C) $ P_{\nu_e\overline{\nu_e}}$, D)$ P_{\nu_e,\overline{\nu_\mu}}$. }
\label{f1}
\end{figure}

\begin{figure}[p]
\centering\hspace{0.8cm}
\epsfig{file=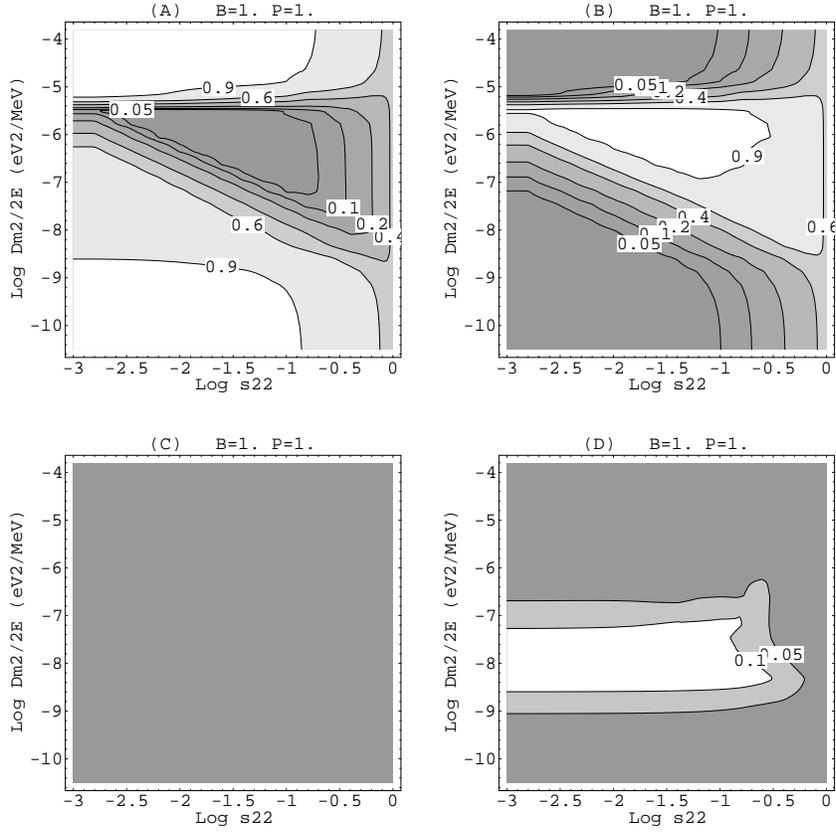,height=16cm}
\caption{As previous figure. Probabilities as a function of $s_2^2$ for 
moderate magnetic field 
($\mu B_{0.7}=1.0\ \mu_{11} B_4\equiv 10^{-11}\mu_B 10^4$ G). }
\label{f3N100}
\end{figure}

In Figs.(\ref{f1}-\ref{f3N100}) we show the electron neutrino survival 
and transition probabilities  for some illustrative cases. 
In Fig.(\ref{f1})  the dependence of the probabilities with $B_{0\perp}$ 
for a fixed small mixing angle ($s_2^2=0.01$) and in absence of 
noisy magnetic field is shown. 
In general spin-flavour precession is suppressed by large mass differences, 
the non-suppression condition is  
$ \mu B >\approx \delta.$
One  can  observe the 
effect of this suppression  around $\delta\approx 10^{-7}-10^{-8}$ eV$^2$/MeV. 
 Magnetic precession 
and MSW transitions respectively 
predominate
 below and above this value. 
For the range
of values considered in this case we observe a very 
modest production of 
$\overline{\nu}_{e}$'s (always 
smaller $<\approx 0.5 \%$ ). The production of
$\overline{\nu}_{\mu}$'s can be however very important. 
In Fig.(\ref{f3N100}) we show the dependence with $s_2^2$  for a fixed magnetic field of very low  magnitude ($\mu B=1.0\ \mu_{11} B_4\equiv 10^{-11} \mu_B \times 10 $ kG).
We observe again the 
same transition region around  
$\delta\approx 10^{-7}-10^{-8}$ eV$^2$/MeV, below this 
range the magnetic precession dominates but for this low value of the 
magnetic field the conversion probability is anyway very small. No significant quantity of either  electron or muon antineutrinos is produced irrespective of the mixing angle.
For stronger magnetic fields 
  (of the order $\mu B\approx 15\ \mu_{11} B_4$) 
the results are similar, the triangular regions, distinctive structure of the 
MSW effect, appear strongly distorted,
the production of muon antineutrinos is notable for 
 $\delta \approx 10^{-6}-10^{-9}$ eV$^2$/MeV
with mixing angles $s_2^2<0.6-0.7$. The production of electron
antineutrinos is not significant even at  these high magnetic fields.

\begin{figure}[p]
\centering\hspace{0.8cm}
\epsfig{file=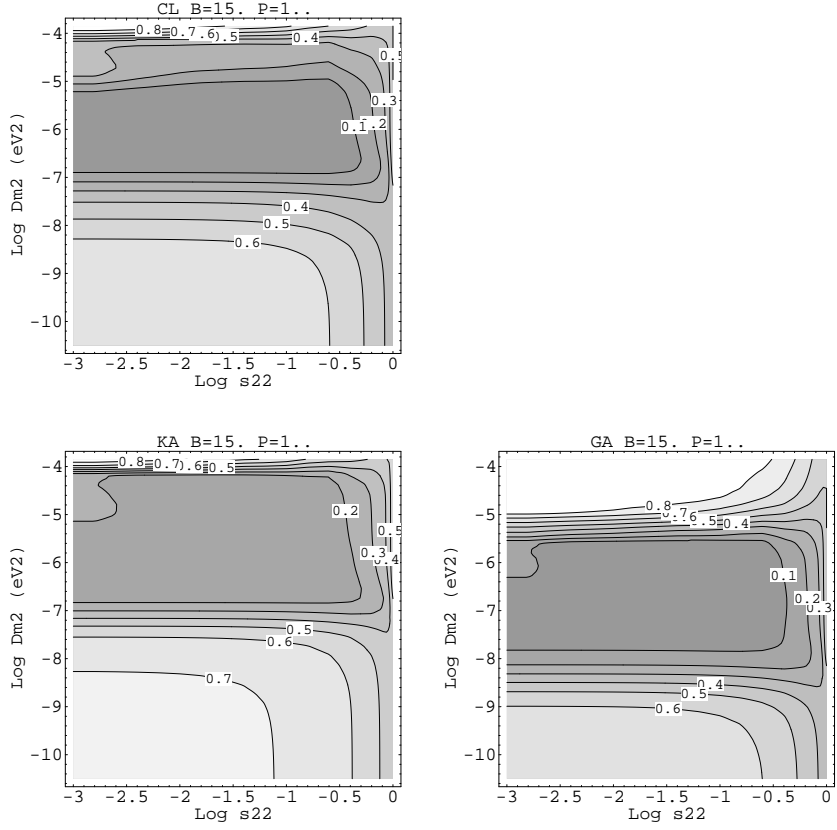,height=16cm}
\caption{Relative signal in absence of noise ($P=1$) 
at the different experiments: (A) Homestake, (B) Kamiokande, (C) Gallium, 
as a function of the mixing angle ($\log s_2^2$, horizontal scale) and squared mass differences 
($\log \Delta m^2$ (eV$^2$), vertical scale). The magnetic field is fixed to be $\mu B=15$ $\mu_{11} B_4$ at the bottom of the convective zone ($r=0.7\ R_\odot$)}
\label{f30N100}
\end{figure}

\begin{figure}[p]
\centering\hspace{0.8cm}
\epsfig{file=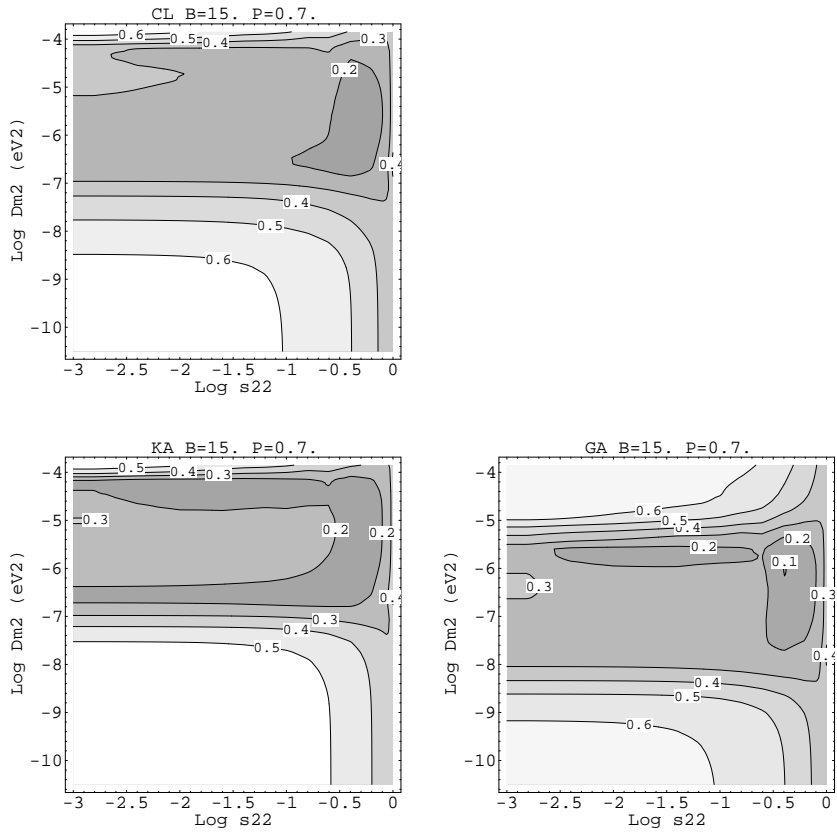,height=16cm}
\caption{The same as Fig.(\protect\ref{f30N100}) but in presence of noise ($P=0.7$, see Table 
\protect\ref{t1}).}
\label{f30N90}
\end{figure}

In the next two figures  we 
show the expected signals in the Chlorine, Gallium and 
Kamiokande experiments for a variety of parameter combinations and fixing  
 the 
transverse magnetic field to a relative high value: $\mu B= 15 \mu_{11} B_4$. 
 Figs.(\ref{f30N100},\ref{f30N90})  correspond respectively to the cases
P=1.00 (absence of magnetic noise), P=0.70 (strong noise). 
The signals in each of the experiments 
have been normalized to the value in absence of neutrino oscillations.
Comparing both figures we can observe in the noisy case how the 
values for the signal approach 1/2. This will be important later when 
we compute the parameter areas consistent with observed total rates, this 
smoothening, particularly in the Ga-Ge experiment will be the responsible
for the appearance of allowed regions at high magnetic fields.

For the Kamiokande experiment we have taken into account 
the four elastic scattering processes $\nu_x e\to \nu_x e$ with 
$\nu_x=\nu_e,\nu_\mu,\overline{\nu}_e,\overline{\nu}_\mu$.
Using the expressions 
for the elastic cross sections 
appearing in \cite{pas1} 
we have defined  an 
``effective'' conversion probability:
\begin{eqnarray}
P_{eff,e}&=& \sum_x c_x
P_{\nu_e \nu_x}\\
c_x&=&\sigma_x (E_\nu,T_{e,min})/\sigma_{\nu_e}(E_{\nu},T_{e,min})
\label{e8050}
\end{eqnarray}
The coefficients $c_x$ 
depend on $T_{e,min}$, the kinetic energy threshold for the observation of the
scattered electron.
For $T_{e,min}\approx 0$ MeV
the coefficients $c_x$ are very accurately independent
of the neutrino energy $E_\nu$ for values
 above  $\approx 1$ MeV. 
For $T_{e,min}\approx 7$ MeV the coefficients $c_x$ associated with
the muon neutrino and antineutrino 
 are still practically 
energy independent while $c_{\overline{\nu}_e}$ 
is only slightly energy dependent.

The effective conversion probability is defined in such a way that
the signal expected at the Super-Kamiokande experiment is:
\begin{eqnarray}
S_{SK}&=&N \int dE\ \epsilon(E) \sigma(E)_{{\nu}_e} \Phi(E) P_{eff,e}(E)
\end{eqnarray}
where $\epsilon$($=1.$ for us) is the experimental detection efficiency, $\sigma_{{\nu}_e}$ the
electron neutrino-neutrino elastic cross section and $\Phi$ the 
total incoming electron neutrino flux predicted by SSM. N is a
normalization constant.

The weighted electron antineutrino 
appearance probability is defined as:
\begin{eqnarray}
\langle P_{\nu_e\overline{\nu}_e}\rangle
&=&\frac{\int_{E>Eth} dE \ \sigma(E) \Phi(E) P_{\nu_e,\overline{\nu}_e}(E)}
{\int_{E>Eth} dE\  \sigma(E) \Phi(E)} 
\label{e4521}
\end{eqnarray}
where $\sigma(E),\Phi(E)$ are respectively the differential cross section 
for the process isotropic background
$\overline{\nu}_e+p\rightarrow e^+ n$ and the differential total neutrino flux coming from the Sun according to
the SSM \cite{BP95}. The threshold energy has been chosen as $E_{th}=9$ MeV. From Kamiokande data we have the
bound $\langle P_{\nu_e\overline{\nu}_e}\rangle
=\approx 0.05$ \cite{fio1,fio2,akh2}.

\begin{figure}[p]
\centering\hspace{0.8cm}
\epsfig{file=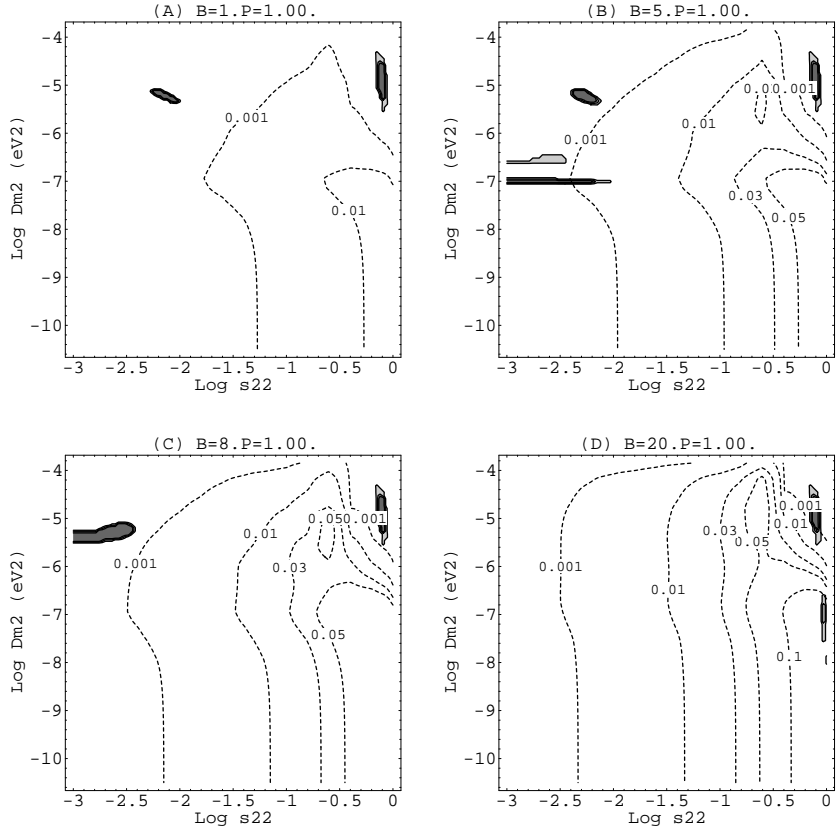,height=16cm}
\caption{The parameter regions consistent with the Homestake, 
Kamiokande and combined Gallium experiments in 
absence of noise: $P=1.00$. 
(C.L.$=90\%,95\%,99\% $ (from darkest to lighter shaded areas)).
Plots (A,B,C,D) correspond respectively to 
values of the regular magnetic field at the 
bottom of the convective zone: $\mu B=1,5,8,20\ \mu_{11} B_{4}$.
The electron antineutrino averaged probability, 
Eq.(\protect\ref{e4521}), is represented by the dashed lines. Present 
Kamiokande bounds impose the additional restriction 
$P_{\nu_e\overline{\nu_e}}<0.05$ to the total rate allowed regions.}
\label{f14N100}
\end{figure}

\begin{figure}[p]
\centering\hspace{0.8cm}
\epsfig{file=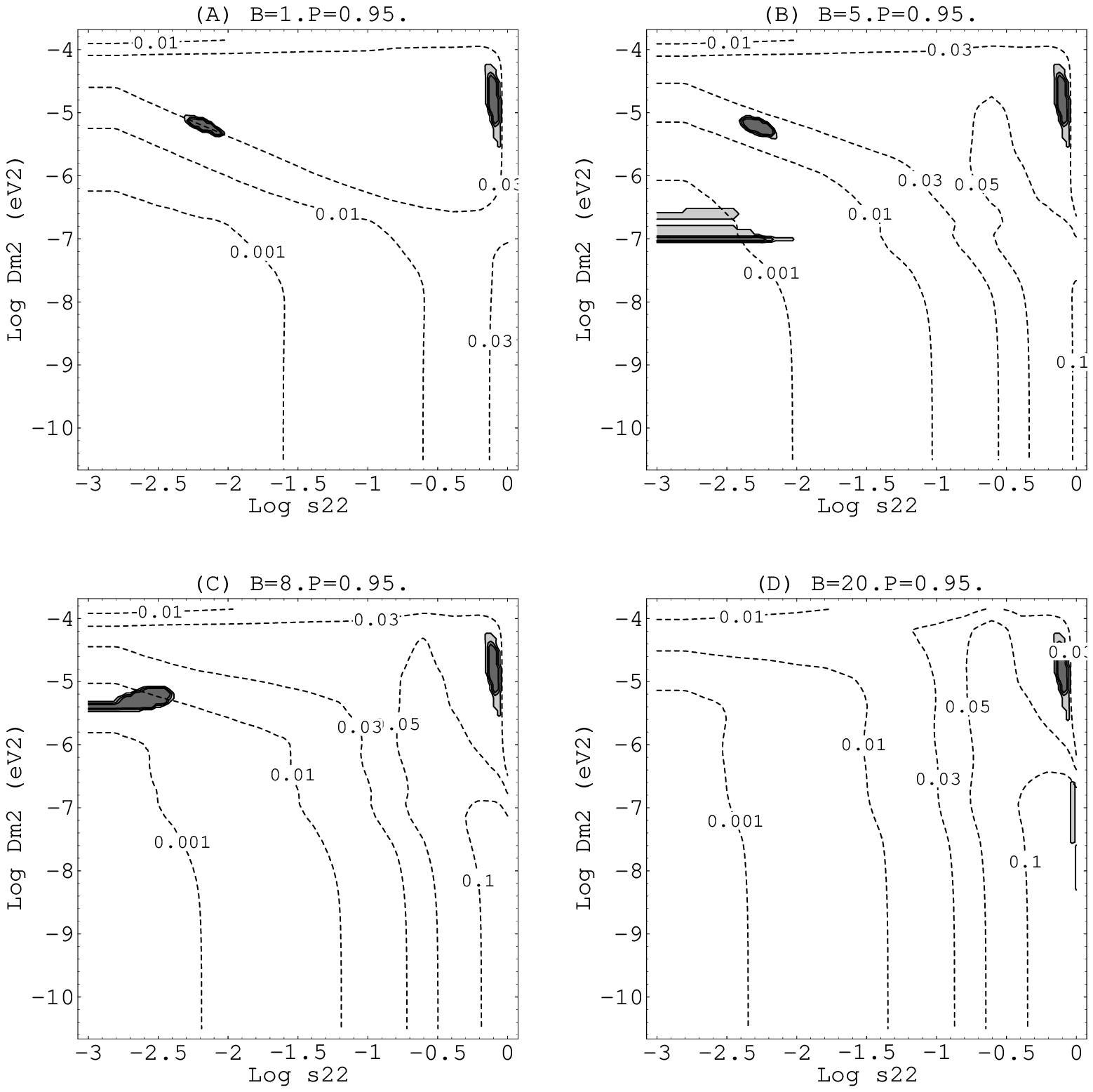,height=16cm}
\caption{The parameter regions consistent with the Homestake, 
Kamiokande and combined Gallium experiments as Fig.(\protect\ref{f14N99}) for $P=0.95$.}
\label{f14N99}
\end{figure}

\begin{figure}[p]
\centering\hspace{0.8cm}
\epsfig{file=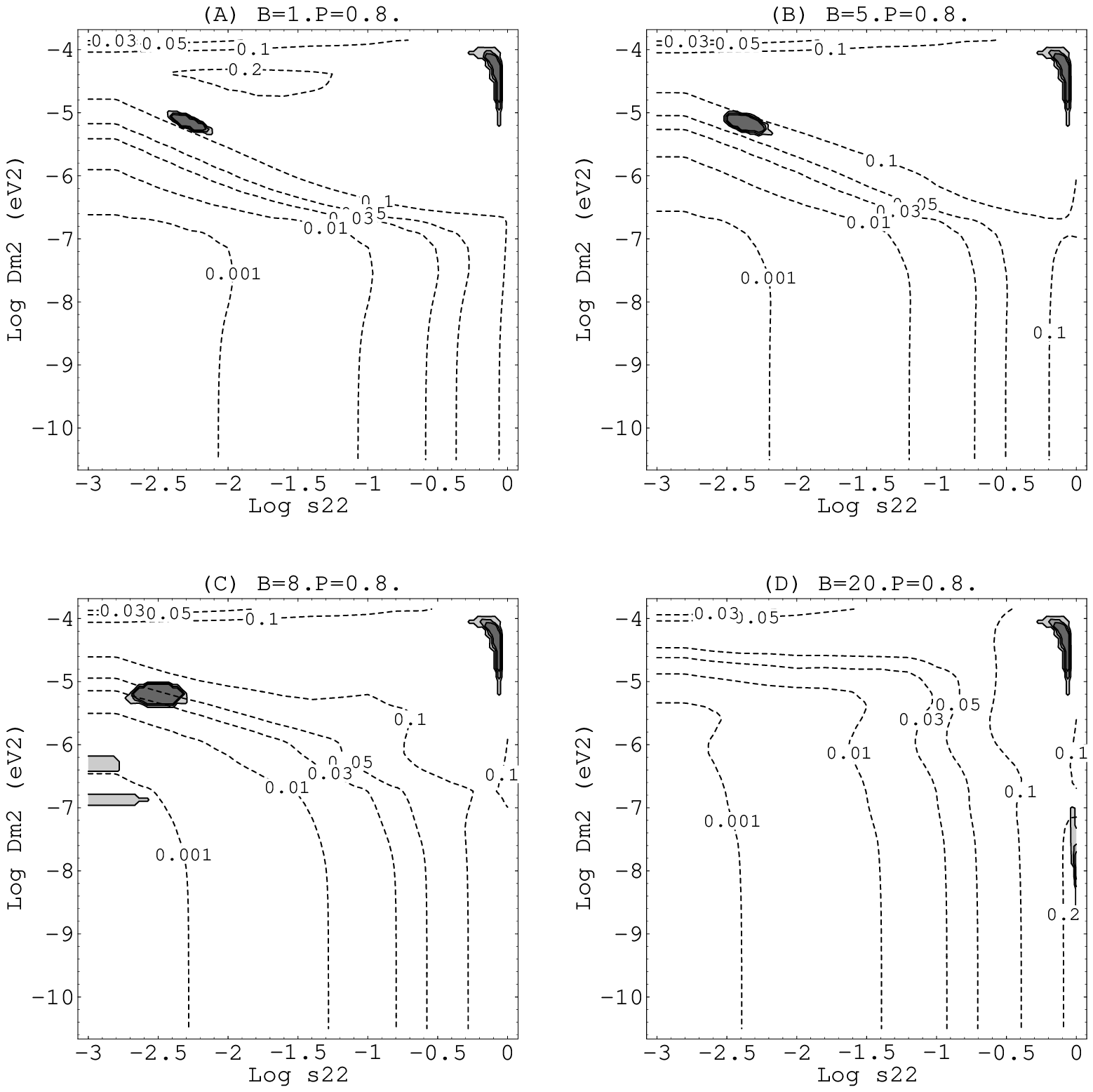,height=16cm}
\caption{The parameter regions consistent with the Homestake, Kamiokande and combined Gallium experiments as Fig.(\protect\ref{f14N100}). for $P=0.80$.}
\label{f14N95}
\end{figure}

\begin{figure}[p]
\centering\hspace{0.8cm}
\epsfig{file=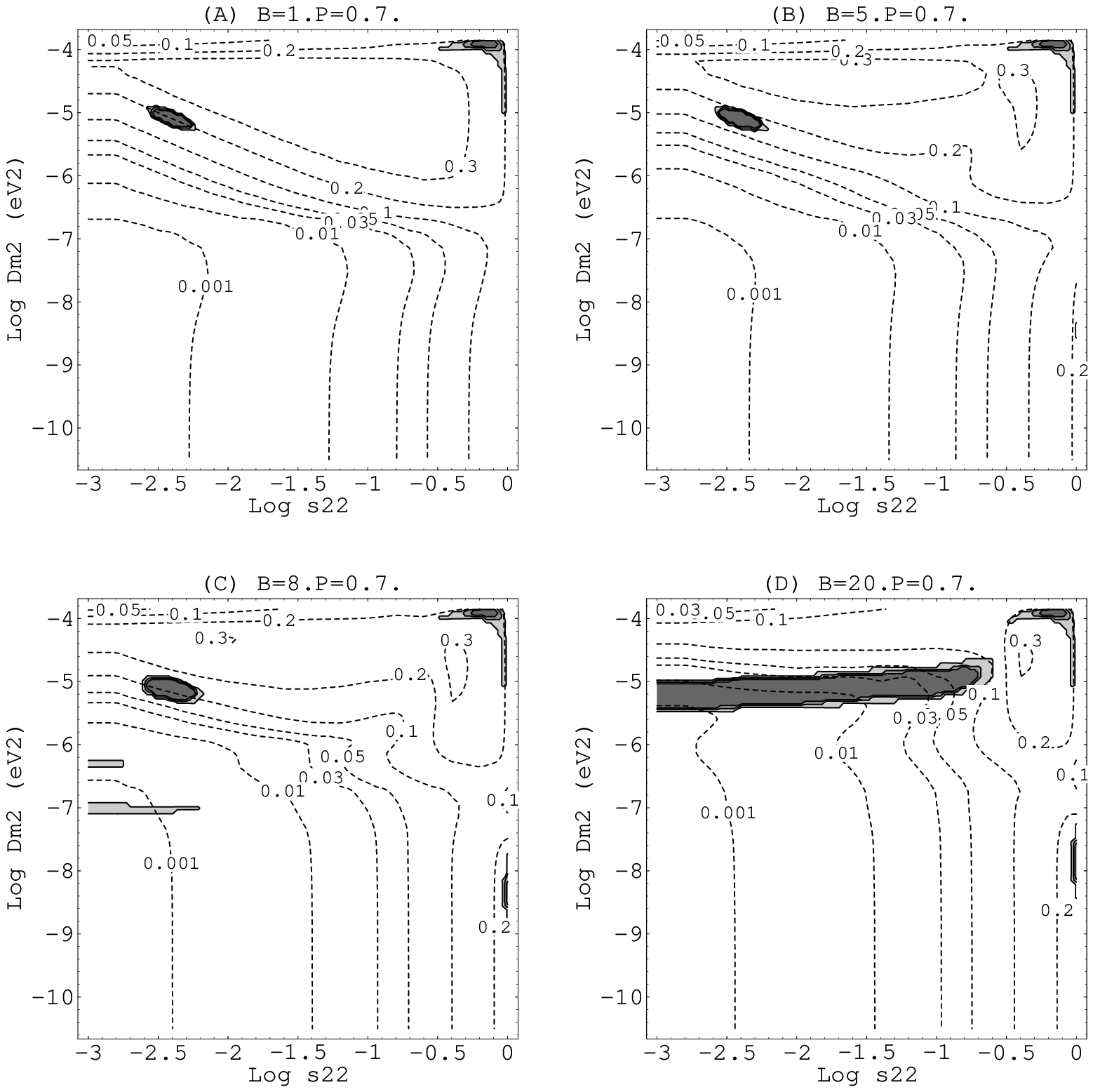,height=16cm}
\caption{The parameter regions consistent with the Homestake, 
Kamiokande and combined Gallium experiments as Fig.(\protect\ref{f14N100}). for $P=0.70$.}
\label{f14N90}
\end{figure}

In Figs.(\ref{f14N100}-\ref{f14N90}) 
we present the 
$(\Delta m^2, \sin^2 2\theta)$ exclusion plots 
from a combined $\chi^2$ analysis of the three experiments
corresponding to the expected signals showed previously.

First we comment the results in complete absence of noise (P=1) which are 
represented
in Fig.(\ref{f14N100}). For negligible or low regular magnetic field we observe
 the  high squared mass difference solutions 
proportioned by the matter MSW effect. As the magnitude of the magnetic 
field increases new solutions appear and disappear in a complicated manner.
The high angle  MSW solution rests practically unmodified for all the 
range considered. The low angle solution however disappears at  high 
magnetic fields, after experiencing some distortion coming from its 
merging with newborn magnetic solutions (compare low angle allowed 
regions in Plots (B) and (C)).
The antineutrino production (dashed lines) is in general  low  and 
Kamiokande bounds are not specially restrictive except at very 
high magnetic fields. Note in Plot (D) the 
very different behavior of the two existing allowed regions:
while the in MSW region the antineutrino production is in the $0.1-1\%$,
 compatible comfortably with Kamiokande bounds, 
the RSFP solution reach a value well above $10\%$ and is excluded by them.  
It seems apparent that there are acceptable particle solutions to the
SNP even for very large regular magnetic fields.

The pattern of the electron antineutrino probability is very different
when  a small level of noise (P=0.95, Fig.(\ref{f14N99})) is switched on.
For $\Delta m^2> 10^{-6}$ eV$^2$: 
the antineutrino iso-probability lines follow 
 the characteristic MSW triangular patterns in this region.
The structure of the allowed regions remain unmodified. The electron 
antineutrino yield in these regions is below the
 $5\%$ level for all cases except in the same region at high mixing angle 
and low squared mass difference as before. 
Note that only low mixing angle solutions with moderate 
regular magnetic fields would be acceptable if
 future data situate the antineutrino bound at the $1-3\%$ level.

\begin{figure}[p]
\centering\hspace{0.8cm}
\epsfig{file=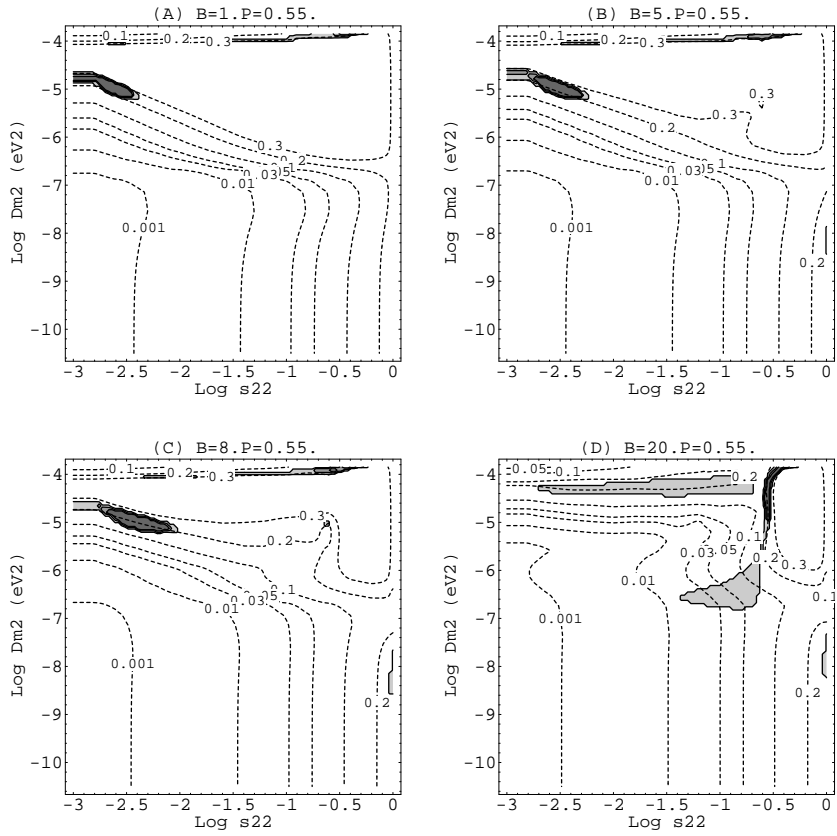,height=16cm}
\caption{$\chi^2$ contour plots.
C.L.$=90\%,95\%,99\% $ (from darkest to lighter). for $P=0.55$.}
\label{f14N80}
\end{figure}

For   stronger levels of noise 
(P=0.8, Figs.(\ref{f14N95}))  the same comments can be said. The structure and 
position of 
the allowed regions from combined total rates are practically
unmodified but the antineutrino yield impose strong restrictions. For 
an antineutrino probability smaller than $3\%$ only some 
 residual, 90\% C.L., allowed regions exist at very small mixing angle, 
$\Delta m^2\approx 10^{-6}-10^{-7}$ eV$^2$ and moderately high 
regular magnetic field (Plot (C)).
The same regions are still acceptable for P=0.7,  
Figs.(\ref{f14N90})). For this level of noise something unexpected 
happens at extremely high 
regular magnetic field (Plot (D)) (probably too high to be acceptable 
on astrophysical grounds): a new, large, 
acceptable region appear for $\Delta m^2\approx 10^{-5}$. 
This region disappears again for extremely 
high chaotic fields (P=0.55, Figs.(\ref{f14N80})).
This apparently erratic behavior could be worthy of a more 
detailed further study.
Note that even for this value of P some residual regions with a 
90\% C.L. are marginally acceptable 
from reconciliation of all experiment total rates and 
antineutrino bounds \cite{fio1}
if the regular magnetic field is $\approx 200$ kG
( for $\mu=10^{-11} \mu_B$).

\section{Conclusions.}

We have presented calculations of neutrino spin flavour precession in 
presence of matter and magnetic field for a theoretically motivated solar
magnetic field profile.
This magnetic field, which is the solution to static magnetic hydrodynamic equations, is a twisting field. 
The degree of twisting and the
 transversal profile are functionally related under this model.
In our calculations we have considered only the solution with minimum
twisting.

Additionally the effect of a layer of magnetic noise at the bottom of the convective zone has been considered, we have justified that the level of noise
in this region can be certainly very high.

We have presented  expected signals and 
expected production of antineutrinos with a without presence of noise. 
We confirm previous results \cite{akh2,akh10} for small mixing and 
 ad-hoc  regular magnetic field profiles.

We find that MSW regions ($\Delta m^2\approx 10^{-5}$ eV$^2$, both small 
and large mixing solutions) are stable up to very large levels of noise (P=0.7-0.8) but they are acceptable from the point of view of antineutrino production 
only for moderate levels of noise ($P\approx 0.95$).

This is in agreement with recent results  obtained through 
a direct ensemble averaging of the solution to  Eq.(\ref{e1001}) \cite{bykov}.
The stronger r.m.s field occurs at the convective zone, the wider 
$(\delta m^2, \sin^2 2\theta)$ region should be excluded when 
considering  the constrain 
imposed by existing antineutrino bounds.

For strong noise, $P=0.7$ or bigger and reasonable regular magnetic 
field, any parameter region 
$(\Delta m^2, \sin^2 2\theta)$ is excluded. This model of noisy magnetic field 
 is not compatible with particle physics solutions to the SNP. One is allowed then  to 
reverse the problem and to put limits on r.m.s field strength, correlation 
length and transition magnetic moments by demanding a solution to the SNP
 under this scenario.

\vspace{1cm}
{\bf Acknowledgments}

The authors thank Sergio Pastor and  Jose Valle for
fruitful discussions. V.B. Semikoz  has been supported by RFFR grant
97-02-16501 and by INTAS grant 96-0659 of the European Union.
E. Torrente-Lujan has been supported by DGICYT under Grant 
 PB95-1077 and by  a DGICYT-MEC contract  at Univ. de Valencia.

\newpage
\begin{table}
\centering
\begin{tabular}{|c|c|c|c|}
 \hline\vspace{0.1cm}
Experiment      & Data~$\pm$(stat.)~$\pm$(syst.)
& Theory  & $S_{Data}/S_{SSM}$
\\[0.1cm] \hline
Homestake       & $ 2.56 \pm 0.16 \pm 0.14$         &
        $7.7^{+1.2}_{-1.0}$     SNU     &  $0.33\pm 0.029  $              \\[0.1cm]
Kamiokande & $ 2.80 \pm 0.19 \pm 0.33$         &
        $5.15^{+1.0}_{-0.7}$  $10^6$ cm$^{-2}$s$^{-1}$ &   $0.54\pm 0.07$  \\[0.1cm]
SAGE            &$66.6^{+7.8}_{-8.1}$&
        $129^{+8}_{-6}$         SNU &                 $    0.52\pm 0.06 $    \\[0.1cm]
GALLEX          & $77.5 \pm  6.2^{+4.3}_{-4.7}$     &
        $129^{+8}_{-6}$         SNU &    $  0.60\pm 0.06 $  \\[0.1cm]
SK (504 days)  &$2.44^{+0.05}_{-0.09}{}^{+0.07}_{-0.06}$&
        $5.15^{+0.93}_{-1.12}$  $10^6$ cm$^{-2}$s$^{-1}$ &$ 0.474\pm 0.020$
\\ \hline       
\end{tabular}
\caption{Neutrino event rates measured by solar neutrino experiments,
        and corresponding predictions from the  SSM
        (see Ref.\protect\cite{bah2}  and references therein, we take the INT normalization for the 
   SSM data).
        The quoted errors are at $1\sigma$.}
\label{t1}
\end{table}

\begin{table}
\centering
\begin{tabular}{|c|c|c|c|c|}
 \hline\vspace{0.1cm}   
P       & $\sqrt{ \langle B_0^2\rangle}$ &
$\eta ,\ (\mu B=1)$& $\eta,\ (\mu B=5)$& $\eta,\ (\mu B=20)$
\\[0.1cm] \hline \vspace{0.0cm}
1.00   & 0.0    & 0.0 & 0.00 & 0.00 \\
0.999  & 2.2    & 5.1  & 0.2 & 0.01 \\
0.95   & 10     & 100  & 4 & 0.2 \\
0.80   & 23     & 540  & 22 & 1.3 \\
0.70   & 34     & 1100 & 46 & 3.0 \\
0.55   & 51     & 2600 & 100 & 6.6 \\
0.51   & 68     & 4600 & 180 & 11 \\
 \hline 
\end{tabular}
\caption{ Values for the noise parameters $\sqrt{\langle B_0^2\rangle}$ and 
$\eta$ (Eq.(\protect\ref{e2003b})) assuming
 $L_0=1000$ km. All $\mu B$ are given in $\mu_{11} B_4$ units.
The quantities $P,\eta$ are dimensionless.}
\label{t2}
\end{table}

\clearpage


\clearpage

\end{document}